\documentstyle[psfig]{lamuphys}
\makeatletter
\let\chapter\hid@chapter
\makeatother
\newcommand{\kms}{{km s$^{-1}$}\/}
\newcommand{\as}{{$''\!\!$.}\/}

\newcommand{\Oiii}{{\sc O~iii}\/}
\newcommand{\Nii}{{\sc N~ii}\/}

\begin{document}
\pagenumbering{arabic}
\title{The Nuclear Disk in M87: A Review}

\author{Holland \,Ford\inst{1,2} and Zlatan\,Tsvetanov\inst{1}}

\institute{Johns Hopkins University, Homewood Campus,
Baltimore, MD 21218, USA
\and
Space Tel. Science Institute,
3700 San Martin Dr., Baltimore, MD 21218, USA}

\maketitle

\begin{abstract}

The disk in the center of M87 is a prototype for gas orbiting a
massive central object.  Three sets of HST+COSTAR FOS and FOC
observations provide strong evidence that the nuclear disk in M87 is
in Keplerian rotation around a black hole with a mass of $(2 - 3)
\times 10^9 M_{\odot}$.  A deep (6 orbits), high resolution
H$\alpha+$[{\sc N ii}] PC2 HST image shows a trailing, three arm
spiral superposed on the underlying nuclear disk.  Several of the
extended filaments appear to connect directly to the disk.  The
filaments extending to the NW appear to be twisted, as in NGC 4258.
Earlier arguments that the NW filaments are flowing from the nucleus
are supported by the presence of blue shifted non-Keplerian components
within 20 pc of the nucleus.  The gas in the blue and red shifted
non-Keplerian components has negative energy and will fall back into
the nucleus.  The morphological and kinematical observations can be
explained by assuming that the filaments originate in a bidirectional
wind emanating from the disk. Such a wind will carry away angular
momentum, enabling gas in the disk to move toward the black hole.

Small ($r \sim 1''; r \sim 100 - 200$ pc), well-defined dusty (D-type)
and ionized (I-type) ``nuclear" disks are common in elliptical
galaxies. We suggest that the size of the black hole's radius of
influence $R_{\rm BH}$ relative to the radius of the nuclear disk
$R_{\rm disk}$ determines whether the disk will be a D-type or I-type.
I-type disks (M87 and M81) occur when $R_{\rm BH} \ge R_{\rm disk}$.
Differential Keplerian rotation throughout the disk may then generate
turbulence and shocks that ionize the gas. D-type disks (NGC 4261 and
NGC 6251) occur when $R_{\rm BH} \ll R_{\rm disk}$. The regions of a
disk that are exterior to $R_{\rm BH}$ will rotate at approximately
constant angular velocity in the galaxy's stellar potential.  In the
absence of differential rotation, shocks will be suppressed, and the
gas will remain cold and dusty. Intermediate D/I types (3C264) may
occur when $R_{\rm BH}$ is a significant fraction of the disk's
radius.  Comparison of $R_{\rm BH}$ with the sizes of the ionized
regions in M87, NGC 4261, and NGC 6251 supports these suggestions.

\end{abstract}

\section{Introduction}

The discovery of a small, well-defined dusty disk in the center of NGC
4261 (Jaffe et al.\ 1994 [J94]; 1996) with a major axis nearly
perpendicular to the galaxy's large scale radio axis inspired the
Faint Object Spectrograph Investigation Definition Team to search for
a similar disk in M87.  H$\alpha$ on-band$/$off-band images taken with
the newly installed WFPC2 (Ford et al.\ 1994) revealed a small ($r
\sim 1''$;~70 pc at 15 Mpc) disk-like structure of ionized gas whose
apparent minor axis was within $\sim 15^{\circ}$ degrees of the
position angle of the jet.  The hypothesis that the gas is a rotating
disk was confirmed when 0\as26 aperture FOS observations made with the
newly installed COSTAR showed Keplerian motion around a central mass
of $2.4 \pm 0.7 \times 10^9 M_{\odot}$ (Harms et al.\ 1994; H94).  The
large mass-to-light ratio $(M/L)_V \sim 500$ of the central mass led
H94 to conclude that the disk is rotating around a massive black hole.
Subsequent observations with the FOS 0\as086 aperture confirmed the
disk's Keplerian motion, but showed that large non-circular motions
are also present (Ford et al.\ 1996a,b; F96).

In this paper we discuss the size, morphology, and alignment of M87's
nuclear disk in the broader context of nuclear disks in AGNs.  We
review the kinematical observations, and suggest a connection between
the non-circular motions and the large scale morphology of the ionized
filaments.  Finally, we present data which supports the hypothesis
that the size of the ionized disk depends on the size of the central
black hole's radius of influence.  Throughout this paper we assume
that the distance to M87 is 15 Mpc.

\section{Nuclear Disks in AGNs: M87}

\index{M87!nuclear disk!classification}
\index{M87!nuclear disk!ionized (I-type)}
\index{M87!nuclear disk!dusty (D-type)}

The disk-like structure in the center of M87 is not unique.  In a
recent review (Ford et al. 1997; F97) we showed that small ($r \sim
1''; r \sim 100 - 200$~pc) well-defined gaseous disks, which we call
nuclear disks, are common in the centers of elliptical galaxies.  The
minor axes of the disks are closely aligned with the directions of the
large-scale radio jets, suggesting that the direction of the jet is
ultimately determined by the angular momentum in the nuclear disk.
The disks are most commonly dusty with unresolved or partially
resolved HII in their centers, rather than completely ionized as in
M87. We will call these D-type disks.  NGC 4261 is the prototype of
the D-type nuclear disks, whereas M87 is the prototype of ionized
nuclear disks, which we call I-type.  Figure 1 gives an example of two
D-type disks (NGC 4261 and NGC 6251) and two I-type disks (M87 and
M81).  There is a striking similarity between the H$\alpha$ ``disk" in
M81 and the disk in M87. However, we do not yet have kinematical
observations of M81 which show that in fact the disk-like structure is
rotating.

\begin{figure}[ht]
\vspace{2.5cm}
\vspace{-50mm}
\centerline{\psfig{figure=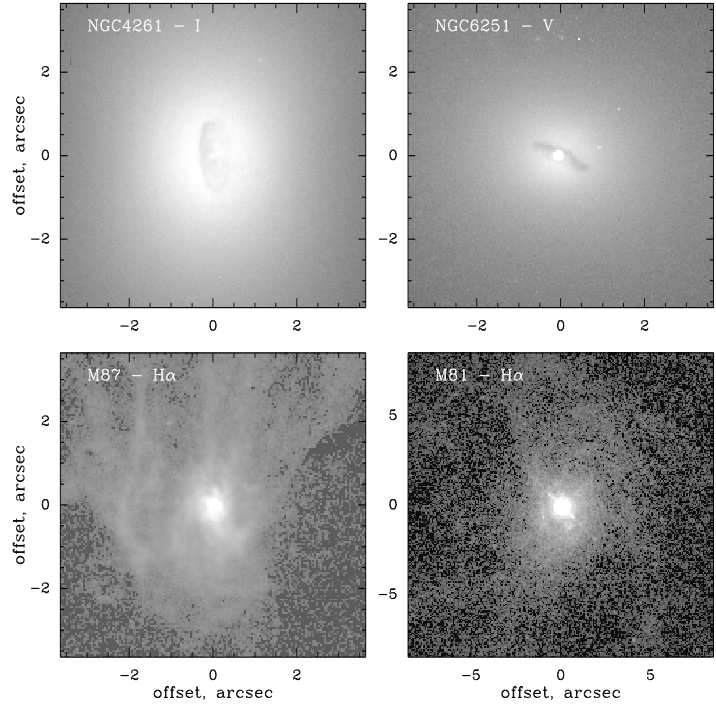,width=11cm}}
\vspace{-22mm}
\caption{Examples of nuclear disks in galaxies with active nuclei.
Two D-type nuclear disks are shown in the upper panels, and two
I-type disks in the lower panels. }
%\vspace{-3mm}
\end{figure}

\index{NGC 4261!black hole mass}
\index{NGC 4261!mass-to-light ratio}

J94 and Ferrarese, Ford \& Jaffe (1996; FFJ96) discussed the
morphological and kinematical evidence for warping in the NGC 4261
disk. FOS 0\as086 aperture observations show that the disk is in
Keplerian motion around a central mass $(4.9 \pm 1.0) \times 10^8
M_{\odot}$ (FFJ96).  The mass-to-light ratio is $(M/L)_V \sim 2100
M_{\odot}/{L_{\odot}}$ within the inner 14.5 pc.  The large $M/L$ and
the fact that NGC 4261 is a radio galaxy strongly point to the dark
mass residing in a black hole.  The angle between the kinematical
minor axis and the radio axis is $21^{\circ}$.

\index{NGC 6251!black hole mass}
\index{NGC 6251!mass-to-light ratio}

The dusty disk in NGC 6251 also is obviously warped.  The fit of the
FOS 0\as086 aperture observations to a Keplerian model gives a central
mass of $(7.5 \pm 2.2) \times 10^8 M_{\odot}$ inside 0\as086 (= 43 pc)
and a central mass-to-light ratio $(M/L)_V \sim
680M_{\odot}/{L_{\odot}}$ (Ferrarese, Ford, \& Jaffe, 1998; FFJ98),
leading to the conclusion that the disk is rotating around a massive
black hole.  The kinematical major axis is rotated $\sim 63^{\circ}$
from the major axis of the dusty disk.  The angle between the
kinematical minor axis and the projection of the radio jet is $\sim
40^{\circ}$.  If the jet is perpendicular to the accretion disk, the
large scale warping of the disk persists down to the accretion disk.

\index{M87!nuclear disk!morphology}

To gain deeper insights into the morphology of M87's H$\alpha$ disk,
and the relationship of the disk to the extended ``filaments" (Ford
and Butcher, 1979; Sparks, Ford, and Kinney 1993; SFK), we used 6 HST
orbits to take deep, dithered, high resolution PC2 F658N observations
of M87.  Figure 2 shows the on-band H$\alpha$ sum of 6 orbits from
which a model of M87's light distribution has been subtracted.  The
H$\alpha$ images are discussed in detail by Tsvetanov et al.\ (1998a)
in this volume.  Here we summarize the principal results.  Out to a
radius of $\sim 1''$ the observed emission-line distribution is well
represented by a trailing three armed spiral superposed on an
elliptical (in projection) power-law disk.  Between 20 and 85 pc the
position angle of the major axis is approximately constant at $\sim
10^{\circ}$, and the ellipticity varies smoothly from 0.2 to 0.4.  The
PA of the jet is $290.5^{\circ}$; the disk minor axis and jet are
aligned to $\sim 10^{\circ}$.  The position angle of the line of nodes
and the mean inclination are close to the values used by H94 and F96
to measure the central mass.

\subsection{Outflow From the Disk}

\index{M87!nuclear disk!outflow}
\index{M87!nuclear disk!filaments}

The filaments in Figure 2 that extend $\sim 17''$ (1200 pc)  to
the NW at position angle $\sim$315$^{\circ}$ appear to be composed of
three distinct twisted strands.  The appearance of these filaments is
similar to the filaments in NGC 4258, which are morphologically (Ford
et al.\ 1986) and kinematically (Cecil, Wilson, \& Tully 1992)
twisted.  The NW filaments at position angles $\sim$315$^{\circ}$ and
$\sim$343$^{\circ}$ appear to connect directly to the disk. Based on
the presence of absorption from dust in the filaments at
$\sim$315$^{\circ}$, SFK argued that these filaments are most likely
tipped toward us, as is the jet.  SFK also found that these filaments
are blue shifted with respect to systemic velocity.  Consequently,
they concluded that the filaments are flowing away from the nucleus,
rather than falling into the nucleus.  There is now direct evidence
for an outflow.  Tsvetanov et al.(1998b; T98) in these Proceedings
found UV absorption lines blue shifted by 150 \kms\ with respect to
M87's systemic velocity.  And, as discussed in Section 4, FOS emission
line spectra show clear evidence for large non-circular motions
associated with the disk.

\begin{figure}
\vspace{-40mm}
\centerline{\psfig{figure=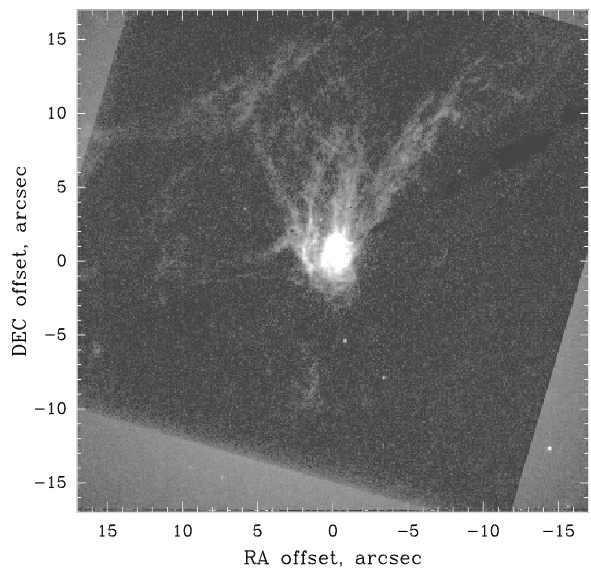,width=11cm}}
\vspace{-27mm}
\caption{The sum of 6 HST orbits of F658N PC2 imaging of M87.  The
starlight has been removed by subtracting a model galaxy.}
\vspace{-3mm}
\end{figure}

We suggest that the filaments which appear to connect to the disk
originate in a wind blowing from the disk.  Because the disk is
rotating, the filaments will carry away angular momentum, allowing gas
to flow through the disk toward the black hole in the center.  The
apparent twisting of the filaments may originate in angular momentum
carried by the wind.  These ideas can be tested by using high spatial
resolution spectroscopy to search for kinematical twisting in the
filaments.

\section{Kinematics and the Central Mass}
\index{M87!nuclear disk!kinematics}
\index{M87!black hole mass}
\index{M87!mass-to-light ratio}

Radial velocities in M87's nuclear disk have been measured with the
Hubble Space Telescope three times.  The first measurements were made
with the FOS by H94 shortly after the installation of COSTAR.
Subsequently the FOS team used FOS observations with 0\as086 and
0\as26 apertures to measure velocities along the disk's minor axis and
within 6 pc of the nucleus along the major axis.  These observations
revealed large non-circular motions at positions along the minor axis
and at some positions close to the nucleus along the major axis.
However, these non-Keplerian velocities appear to be superposed on
Keplerian rotation around a massive black hole (F96).  Finally,
Macchetto et al. (1997; M97) used COSTAR and the FOC 0\as06 $\times$
13\as5 slit to measure [{\sc O~ii}] $\lambda\lambda$3727+29 velocities
across the disk.  Although their difficult target acquisition missed
the nucleus, their data is a good fit to Keplerian rotation around a
mass of $3.2 \pm 0.9 \times 10^9 M_{\odot}$.

Table 1 summarizes the results from the three sets of observations.
The F96 analysis includes the H94 data, so the two mass determinations
are not independent.  The central mass derived from the FOS and FOC
observations agree to within their respective error bars.  The
conclusion that the disk is in rapid rotation around a central mass of
$(2-3) \times 10^9 M_{\odot}$ appears to be firm.

The gas in the disk is very turbulent; 600 \kms\ is the characteristic
FWHM.  The energy in this turbulence has not been included in the
measurements of the central mass made by H94, F96, and M97. At
positions 9a,b and 11a,b (cf Section 4)the circular velocity in the
disk is $\sim 1200$ \kms\ for a mass of $2 \times 10^9 M_{\odot}$.  We
take $3 {\sigma}^2 / v_{\rm circ} \sim 0.13$ as a measure of how much
the mass has been underestimated. Neglecting the turbulence does not
appear to have much effect on the estimated mass.  This reflects the
fact that in spite of the turbulence, the disk does appear to be in
Keplerian rotation.

%\vspace{-5mm}
\begin{table}
\begin{center}
\caption{HST Spectroscopic Observations of the Nuclear Disk in M87}
\begin{tabular}{lccccc}
\hline
Paper & Instrument & No. of Positions & Mass &Radius & $(M/L)_{V}$ \\
      &            &   & ($M_{\odot}$) & (pc) & $(M/L)_{\odot}$ \\
\hline\\[-9pt]
H94 &FOS$+$COSTAR &$5 \times 0$\as26 & $2.4\pm 0.7 \times 10^9	$ &18& $\sim$ 540 (starlight)\\
F96 &FOS$+$COSTAR&$6 \times 0$\as086 & $2.0 \pm 1.0 \times 10^9 $ &$\sim$6 & $\sim$3100 (starlight)\\
&&		$2\times 0$\as26&&&\\
M97 & FOC$+$COSTAR &  3$\times$0\as06$\times$13\as5 & $3.2 \pm 0.9 \times 10^9$ & $\sim$6	& $\sim$ 110 (total light)\\
\hline
\end{tabular}
\end{center}
\end{table}

\index{M87!dark central cluster}
F96's mass-to-light ratio $(M/L)_V \sim 3100~(M/L)_{\odot}$ excludes
the non-thermal light from the unresolved point source (cf T98),
whereas the value reported by M97 is based on the total light.  An
$M/L$ that is orders of magnitude larger than the $M/L$ found in star
clusters of similar color is compelling, but not conclusive, evidence
that the dark mass resides in a black hole.  Maoz (1995, 1998) has
investigated the maximum lifetimes of central dark clusters composed
of compact objects such as white dwarfs, neutron stars, and stellar
mass black holes.  He finds that present observations impose maximum
ages of central dark clusters that are much less than the age of the
galaxies in only two galaxies, the Milky Way and NGC 4258.  Unless the
dense objects in the hypothetical nuclear clusters have masses $\leq
0.03 M_{\odot}$ (e.g, low-mass black holes or elementary particles),
the respective maximum ages of dark clusters are $\sim 10^8$ yr and
$\sim 2 \times 10^8$ yr.  Because these lifetimes are much less than
the ages of galaxies, dark clusters are highly improbable sources for
the mass.  Consequently, the Milky Way and NGC 4258 are the strongest
candidates for hosting massive black holes.

In spite of the theoretical possibility that a dark cluster may
provide M87's dark mass, we think there are many reasons for
concluding that the dark mass resides in a black hole.  These
arguments, which are largely encapsulated within the ``AGN Paradigm,"
include large energy release in the small volumes implied by
variability, and the production of relativistic jets from the presumed
accretion disks around a massive black hole.  Given the high
probability that massive black holes exist in two AGNs, we think it
likely that the large dark masses routinely being found in AGNs such
as M87, NGC 4261, and NGC 6251 (FFJ98) also reside in massive black
holes.

\section{Non-Keplerian Motions}

\index{M87!nuclear disk!kinematics}
\index{M87!nuclear disk!non-Keplerian motions}
Non-Keplerian motions are readily evident when non-systemic velocities
are observed at positions along the minor axis of the disk, or when
the line profiles show the unmistakable presence of two or more
emission line components.  Because of the difficulty of separating
individual velocity systems in the H$\alpha + $[\Nii] blend, we used
the [\Oiii]~$\lambda$5007 profiles to identify the non-Keplerian
components.  Figure 3 is a schematic that shows the positions on the
major and minor axes where we find non-circular motions.  Figure 4
shows the spectra of H$\beta$ through [\Oiii] at position 7 and 8 on
the minor axis.

At position 7 on the minor axis, which is on the same side as the jet,
the systemic component of the disk is present at $\sim 1300$~\kms\ and
$\sim 1193$~\kms\ in [\Oiii]~$\lambda$5007 and H$\beta$.  A blue
shifted component is clearly present in both [\Oiii]~$\lambda$5007 and
H$\beta$ at respectively $675$~\kms\ and $544$~\kms.  Note that the
blue shifted component is stronger than the systemic component in
[\Oiii], whereas the reverse is true in H$\beta$.
[\Oiii]~$\lambda$4959 has the same profile as [\Oiii]~$\lambda$5007,
although the weaker systemic component is present only as an
inflection on the red side of the line profile.

At position 8 on the minor axis opposite the jet, the expected
systemic component at $v \sim 1325$~\kms\ is present, as well as a red
shifted component at $v \sim 1775$~\kms and a {\it blue} shifted
component at $v \sim 884$~\kms\~.  The line [\Oiii]~$\lambda$4959 has
the same profile as $\lambda$5007, so we are confident that all three
components are present.  H$\beta$ is very broad, undoubtedly due to
the presence of multiple components.  As at position 7, [\Oiii] and
H$\beta$ have different profiles, showing that the excitation varies
between the different components.

\begin{figure}[ht]
%\vspace{2.5cm}
\vspace{-27mm}
\centerline{\psfig{figure=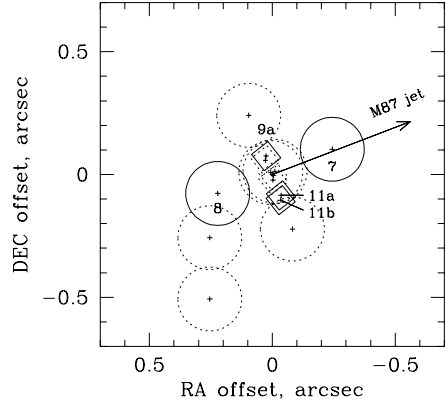,width=10cm}}
\vspace{-20mm}
\caption{A schematic showing the positions where FOS measurements were
made.  The position of the nucleus is at the base of the arrow
representing the jet. Solid lines show the positions where non
Keplerian velocities were found. }
\vspace{-3mm} 
\end{figure}

\begin{figure}[ht]
%\vspace{2.5cm}
\vspace{-2mm}
\centerline{\psfig{figure=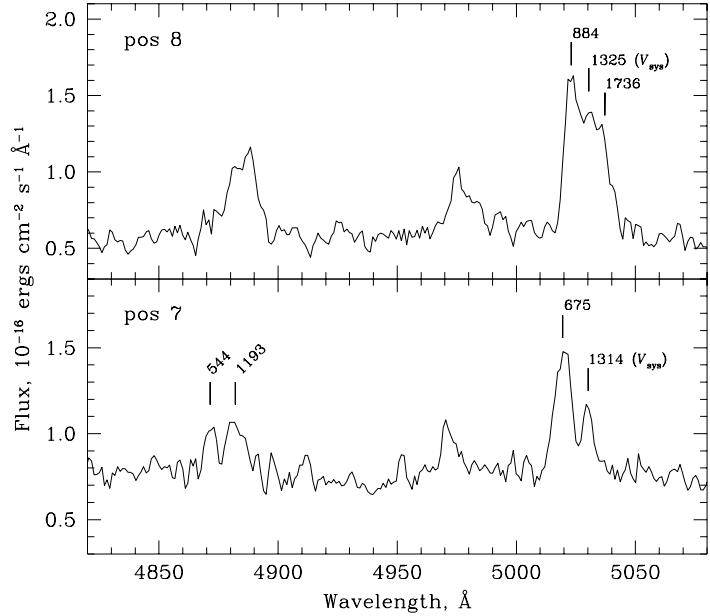,width=10cm}}
\vspace{-15mm}
\caption{The upper panel shows the nuclear spectrum from H$\beta$
through [\Oiii] $\lambda$5007 at position 8 on the minor axis opposite
the jet.  The systemic component from the disk is flanked by blue
shifted and red shifted components.  The lower panel shows the
spectrum at position 7 on the minor axis on the same side as the jet.
The blue shifted component is obvious in both [\Oiii] $\lambda$5007
and H$\beta$.}
\vspace{-3mm} 
\end{figure}

\vspace*{-8mm}
\begin{table}
\begin{center}
\caption{Non-Keplerian Velocities Within 0\as26 of the M87 Nucleus}
\begin{tabular}{lrrrrrc}
\hline
Position&$x\phantom{??}$&$y\phantom{??}$&$~v_{\rm obs}\phantom{??}$& $~~v_{\rm k}\phantom{??}$~~&\phantom{???}$r_{0}\phantom{??}$&\multicolumn{1}{c}{$v_{\rm w}/v_{\rm circ}$}\\
&($''$)\phantom{?}&($''$)\phantom{?}&~~~km s$^{-1}$&\phantom{??} km s$^{-1}$&(pc)\phantom{?}&\\
\hline
9a  &0.0250		&0.0750	& 766\phantom{??} 	&1984\phantom{??} &5.5\phantom{?} 	& 0.48\\
8   &0.2220		&$-$0.0770	&1736\phantom{??} 	&1232\phantom{??} & 20.1\phantom{?} & 0.92\\
    &0.2220		&$-$0.0770	& 884\phantom{??} 	&1232\phantom{??} &20.1\phantom{?} 	& 0.69\\
7   &$-$0.2440	&0.1030	& 675\phantom{??} 	&1287\phantom{??}	& 22.7\phantom{?}	& 1.15\\
11a &$-$0.0355	&$-$0.0845	& 851\phantom{??} 	&576\phantom{??} 	&6.4\phantom{?}	& 0.42\\
    &$-$0.0355	&$-$0.0845	&1424\phantom{??}  &576\phantom{??} 	& 6.4\phantom{?} 	& 0.18\\
11b &$-$0.0345	&$-$0.1045	&1503\phantom{??}  &629\phantom{??} 	& 7.6\phantom{?} 	& 0.29\\
\hline
\end{tabular}
\end{center}
\end{table}

The observed non-Keplerian components are summarized in Table 2.  The
first column gives the position designations assigned after the
observations were made (e.g. Position 11a was the first visit to
Position 11, and Position 11b was the second visit).  Columns 2 and 3
give the locations of the apertures relative to the nucleus derived
from the FOS peak-up data ($x$ and $y$ are positive to the E and
N). Column 4 gives the observed velocities of the non-Keplerian
components.  Column 5 gives the {\it predicted} line-of-sight
velocities from the model disk (F96), calculated using the equations
in H94.  Column 6 lists the radius in the disk at the position of the
aperture.  Finally, column 7 gives the ``wind" velocity, $v_{\rm w} =
(v_{\rm obs} - 1250) / cos(i)$, divided by the circular velocity
$v_{\rm circ}$ at the position $r_0$, assuming that the wind is
perpendicular to the disk.

The blue shifted velocity component (675 \kms) at position 7 along the
minor axis on the side of the jet could be due to gas entrained by the
jet.  There is a corresponding red shifted component (1736 \kms) at
position 8 on the minor axis opposite the jet.  This could be gas
entrained by the unseen counter jet.  However, there also is a {\it
blue} shifted component (884 \kms) at position 8.  Further more, there
are both red and blue shifted components at positions 9a, 11a, and 11b
along the major axis on both sides of the nucleus.  We think the most
likely explanation for these components is that a ``wind" is coming
off the disk.  As previously noted, SFK argued that the filaments
extending to the NW are an outflow from the nucleus.  Because these
filaments appear to connect to the disk, it is natural to suppose that
they originate in a ``wind" that is carrying away angular momentum.
Although we cannot directly associate the non-Keplerian velocity
components with these filaments, we think they lend support to this
suggestion.

Escape from the black hole's sphere of influence requires $v_{\rm w}
\ge \sqrt{2}~v_{\rm circ}$.  Unless the gas is far from the disk,
Table 2 shows that the gas is bound to the black hole, and will fall
back in.  This may account for the apparent ``loop" structure which
extends $\sim 8''$ to the NNW of the nucleus (see Fig.~2).  The fact
that the gas is bound to the nucleus should not be surprising.  Unless
there is a large unobserved reservoir of energy in the nuclear disk,
such as a strong magnetic field, it will be impossible to drive off
gas at velocities exceeding escape velocity.

\section{Disk Type and the Black Hole's Radius of Influence}

\index{nuclear disks!classification}
NGC 4261 and NGC 6251 are examples of well defined, dusty (D-type)
disks that have small, partially resolved regions of ionized gas in
their centers. Conversely, the disk in M87 is entirely ionized
(I-type). The face-on disk in 3C 264, shown in Figure 5, is
intermediate between NGC 4261 and M87.  Disks such as this one,
wherein the ionized region is more than $\sim 20\%$ of the radius of
the dusty region, will be referred to as D/I-type.

\begin{figure}[h]
%\vspace{2.5cm}
\vspace{-25mm}
\centerline{\psfig{figure=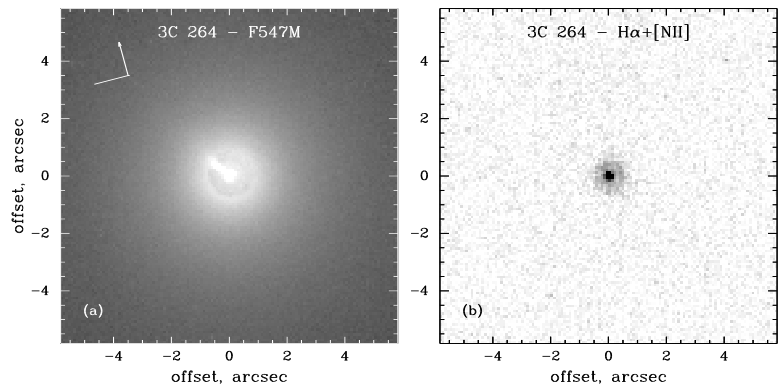,angle=-90,width=12cm}}
\vspace{-15mm}
\caption{The left panel shows a PC2 F547M continuum image of 3C264.
The bright linear feature in the F547M continuum image of 3C264 that
projects to just beyond the edge of the disk is an optical jet (cf
Baum et al.\ 1997).  The right hand panel shows an H$\alpha$+[\Nii]
image obtained by subtracting the F547M image from a F673N image. A
small disk of ionized gas is present in the center of the dusty disk.}
\vspace{-3mm} 
\end{figure}

What determines the type of nuclear disk? We hypothesize that the type
of disk is determined by the size of the disk relative to the black
hole's radius of influence.  If the dispersion of stars in the bulge
of the parent galaxy is $\sigma_0$, the black hole's radius of
influence will be

		\begin{equation}				 
		{\rm R_{\rm BH} \sim GM_{\rm BH} /{\sigma_0}^2 = 
		43~M_8/(\sigma_{100})^2~{\rm pc}}
		\end{equation}

\noindent where $\sigma_{100} = \sigma_0/100$ km s$^{-1}$, and ${\rm
M_8 = M_{\rm BH}/10^8 M_{\odot}}$.  We suggest that when $R_{\rm disk}
\le R_{\rm BH}$, differential Keplerian rotation generates turbulence
and shocks throughout the disk that ionize the gas.  H94 found that
the gas in the disk is very turbulent with a characteristic FWHM
$\sim$ 500 \kms.  Dopita et al.\ (1997) and Dopita (1998) find that
the gas in the disk is ionized by fast shocks rather than by nuclear
photoionization.  Shock ionization is consistent with the presence of
turbulence and the spiral features in the disk.

The regions of a disk that are exterior to the black hole's radius of
influence will rotate at an approximately constant angular velocity in
the galaxy's stellar potential.  In the absence of differential
rotation, shocks will be suppressed, and the gas will remain cold and
dusty.

Table 3 lists the three galaxies to date that have nuclear disks and a
measured black hole mass.  The radius $R_{\rm BH}$ was calculated with
the assumption that all three galaxies have a nuclear stellar velocity
dispersion $\sigma_0 = 300$ \kms.  The fact that the disk in M87 is
I-type, whereas the disks in NGC 4261 and NGC 6251 are D-type, is
consistent with our hypothesis.

\vspace*{-5mm}
\begin{table}
\begin{center}
\caption{Disk Radius and the Black Hole's Radius of Influence}
\begin{tabular}{l@{\tabskip.35truein}cccc}
\hline
Galaxy & $M_{\rm BH}$ & $R_{\rm BH}$ & $R_{\rm Disk}$ & Disk Type\\
       & $M_{\odot}$  & (pc)         & (pc) & \\[2pt] \hline \\[-9pt]
M87   & $(2-3)\times 10^9$ & 100--140 & $\sim$100 & I \\
N6251 & $7.5 \times 10^8$  & 36       & 330       & D \\
N4261 & $4.9 \times 10^8$    & 19       & 130       & D \\
\hline
\end{tabular}
\end{center}
\end{table}
\vspace*{-5mm}

\section{Summary} 

The disk in the center of M87 is not unique.  Small ($r \sim 1''; r
\sim 100 - 200$ pc), well-defined dusty (D-type) and ionized (I-type)
``nuclear" disks are common in elliptical galaxies (F97). The minor
axes of these disks are closely aligned with the directions of the
large scale radio jets, suggesting that the disk's angular momentum
determines the direction of the radio jets.

HST observations of M87 with the FOS+COSTAR (H94 and F96) and the
FOC+COSTAR (M97) provide strong evidence that the I-type nuclear disk
in M87 is in Keplerian rotation around a central mass of $(2-3) \times
10^9 M_{\odot}$.  The stellar $(M/L)_V$ at radii $r \le 6$ pc is at
least 3100 $(M/L)_{\odot}$, a value orders of magnitude larger than
found in star clusters.  The high $M/L$ combined with i) arguments
encapsulated in the ``AGN Paradigm,'' and ii) the nearly inescapable
fact that there are massive black holes in the centers of the Milky
Way and NGC 4258, strongly suggest there is a massive black hole in
M87.

Deep, high resolution HST images show a three arm spiral superposed on
the underlying nuclear disk .  The filaments extending to the NW
appear to be twisted, as are the filaments in NGC 4258.  Several of
the filaments appear to connect directly to the disk.  SFK's arguments
that the NW filaments are flowing out from the nucleus are supported
by the presence of blue shifted non-Keplerian components seen in
absorption and in emission.  The non-Keplerian components do not have
enough kinetic energy to escape the black hole's potential, and will
fall back into the nucleus.  The morphological and kinematical
observations can be explained by assuming that the filaments originate
in a bidirectional wind that is blowing off the disk.  If this is
true, the wind removes angular momentum and allows the gas in the disk
to move toward the black hole.

The radius of the central black hole's radius of influence relative to
the size of the nuclear disk determines whether the disk will be a
dusty D-type or an ionized I-type disk.  D-type disks (NGC 4261 and
NGC 6251) occur when the black hole's radius of influence is much
smaller than the disk's radius.  I-type disks (M87 and M81) occur when
the radius of influence is equal to or greater than the radius of the
nuclear disk.  Intermediate D/I types (3C264) occur when the radius of
influence is a significant fraction of the disk's radius.

%
% ---- Bibliography ----
%

\end{document}